\documentclass[journal=jpccck,manuscript=article,layout=twocolumn]{achemso}

\setkeys{acs}{
abbreviations=true,
articletitle=true,
biblabel=plain,
chaptertitle=true,
doi=true,
email=true,
etalmode=truncate,
keywords=true,
maxauthors=100,
super=true
}

\setcitestyle{super,open={},close={}}

\usepackage[utf8]{inputenc} 
\usepackage{textcomp} 
\usepackage[T1]{fontenc} 
\usepackage[portuguese,english]{babel} 
\usepackage[version=4]{mhchem} 
\usepackage{siunitx} 
\usepackage{graphicx} 
\graphicspath{{Figures/}} 
\usepackage{geometry} 
\usepackage[format=plain,justification=justified,singlelinecheck=false,font={normalsize,bf},labelfont=bf,labelsep=space]{caption} 
\usepackage{float} 
\usepackage{natbib} 
\usepackage{setspace} 
\usepackage{xkeyval} 
\usepackage{array} 
\usepackage{booktabs} 
\usepackage{listings} 
\usepackage{lmodern} 
\usepackage{mathpazo} 
\usepackage[breaklinks,colorlinks=true,allcolors=blue]{hyperref} 
\hypersetup{breaklinks,colorlinks=true,linkcolor=blue,filecolor=blue,urlcolor=blue,citecolor=blue}
\usepackage{subfigure} 
\usepackage[compact]{titlesec} 
\usepackage{natmove} 
\usepackage{multirow} 
\usepackage{lipsum} 
\usepackage{calc} 
\usepackage{longtable} 
\usepackage{tabularx} 
\usepackage{placeins} 
\usepackage{tablefootnote}
\usepackage{adjustbox}
\usepackage{titletoc}
\usepackage{latexsym}
\usepackage{cancel}
\usepackage{amsmath}
\usepackage{amssymb}
\usepackage{amsfonts}
\usepackage{amstext}
\usepackage{cuted}
\usepackage{braket}

\usepackage{tikz}
\usepackage{stfloats}
\usetikzlibrary{positioning}

\usepackage[none]{hyphenat}
\usepackage{microtype}
\makeatletter
\let\l@addto@macro\relax
\makeatother
\usepackage[fontsize=12pt]{scrextend}

\SectionsOn
\SectionNumbersOn
\AbstractOn

\title[Short Title]{Geometry-Induced Effective Tight-Binding Hamiltonians for Phononic Systems: Mode Conversion and SSH-Like Physics}

\author{José E. González}
\affiliation{Departamento de Física, Facultad de Ciencias, Universidad Nacional Autónoma de México, Ciudad de México 04510, México}

\author{Carlos Ramírez}
\email{carlos@ciencias.unam.mx}
\affiliation{Departamento de Física, Facultad de Ciencias, Universidad Nacional Autónoma de México, Ciudad de México 04510, México}

\abbreviations{}


\begin{document}

\maketitle


\begin{abstract} 
We present a framework that maps harmonic vibrational systems onto effective multi-orbital tight-binding Hamiltonians, establishing a direct correspondence between phononic degrees of freedom and graph-based lattice models. Within this mapping, the Cartesian displacement components of each mass become internal orbitals, while elastic interactions generate effective onsite energies and hopping amplitudes determined by the geometry of the system. This representation enables the application of numerical and conceptual tools originally developed for electronic transport to the study of phononic systems. The method is first applied to a monoatomic chain containing an angular bend. For this benchmark system, analytical expressions for the polarization-resolved transmission and reflection probabilities are derived and compared against calculations performed using a recursive scattering-matrix method within the mapped tight-binding representation. Excellent agreement is obtained, validating the mapping and demonstrating its ability to describe geometry-induced mode conversion and the influence of evanescent states. We then investigate zigzag chains, where alternating bond orientations generate effective dimerized Hamiltonians reminiscent of the Su--Schrieffer--Heeger (SSH) model. Although the uniform zigzag chain develops a spectral gap, no localized edge states are observed. We show that modified boundary parameters generated by the phononic mapping suppress the edge-state formation expected from the ideal SSH picture. Introducing a geometric domain wall restores localized domain-wall states inside the gap, which give rise to resonant transmission channels across an otherwise insulating frequency window. The resonance energies remain robust against system-size variations and the corresponding transmission approaches unity. These results demonstrate that geometry can generate effective tight-binding structures with nontrivial localization and transport properties in phononic systems, providing a versatile framework for the analysis and design of complex vibrational networks. 
\end{abstract}

\section{Introduction}
Understanding vibrational properties in lattice systems is central to condensed matter physics, with direct implications for thermal transport, mechanical response, and the design of phononic devices \cite{Ashcroft,Kittel,Nomura,Xie}. In crystalline solids, lattice vibrations are described in terms of normal modes obtained from the diagonalization of the dynamical matrix, which encodes the harmonic interactions between degrees of freedom \cite{Kittel,Born}. This formalism is particularly well suited to periodic systems or configurations that admit analytical simplifications based on symmetry \cite{Ashcroft}.

However, in many physically relevant situations—such as systems with geometric defects, curved structures, or spatially varying orientations—the analysis of phonon propagation becomes more involved \cite{Rabia,Hao}. Geometric features can induce coupling between longitudinal and transverse modes, leading to mode conversion and nontrivial transport phenomena that are difficult to capture within standard approaches \cite{Lee,Long,Cao,DalPoggetto,Chen,Krushynska}. These considerations motivate the development of alternative frameworks capable of describing phonon dynamics in complex geometries while maintaining a clear connection to the underlying physical mechanisms \cite{Beardo,Yamamoto,Chen2,Barreto_2025}.

In parallel, the study of electronic transport in condensed matter systems has greatly benefited from tight-binding formulations \cite{Datta} and their associated graph representations, in which lattice sites and hopping processes provide an intuitive description of quantum dynamics that can be readily adapted to complex geometries \cite{Klimeck,Karamlou}. These approaches naturally lend themselves to transport calculations through techniques such as Landauer--B\"uttiker formalism \cite{Buttiker} and scattering \cite{Waintal} or Green's function methods \cite{Rotter}, which are well suited to systems with disorder, interfaces, and nontrivial connectivity \cite{Rodriguez}. Despite formal similarities between the eigenvalue problems governing electronic and vibrational systems, tight-binding perspectives are not typically adopted as a general framework for the analysis of phonons, particularly in systems with complex geometrical constraints where they could offer a natural and efficient description.

In this work, we develop a framework that systematically maps lattice vibrational systems onto effective tight-binding models with multiple internal degrees of freedom per site. Within this mapping, the spatial displacements of each mass are identified with internal orbitals in a multi-orbital lattice model, while force constants define effective hopping amplitudes and onsite terms. This construction leads to a graph representation of the phononic system, in which geometry is encoded explicitly in the connectivity and strength of the effective couplings. The approach provides a systematic procedure to build tight-binding graphs from arbitrary elastic networks, enabling the application of analytical and numerical techniques originally developed for electronic transport to phononic systems. Furthermore, the mapping reveals effective lattice structures whose spectral and transport properties are directly controlled by geometry.

Within this framework, geometric misalignment between neighboring elements gives rise to coupling between different vibrational polarizations. In the tight-binding language, this appears as off-diagonal hopping terms between internal orbitals, leading to hybridization controlled by the underlying geometry. This mechanism corresponds to a geometrically induced mixing of longitudinal and transverse modes and highlights the role of geometry in generating effective couplings.

To illustrate this approach, we first analyze a minimal building block consisting of two masses coupled by anisotropic elastic interactions at an arbitrary relative angle. From this elementary unit, we derive the corresponding tight-binding graph and identify the effective hopping structure responsible for mode mixing. We then apply the framework to investigate phonon transport in two representative systems. In the first case, we consider a one-dimensional chain that undergoes a sudden change in direction, introducing a localized geometric defect. Transmission and reflection probabilities are computed both analytically, using conventional phonon techniques, and numerically via a recursive scattering matrix method \cite{Rodriguez} formulated in the tight-binding mapped representation. The comparison between these approaches supports the validity of the mapping and reveals the role of geometry in inducing mode conversion.

In the second case, we examine a zigzag chain in which the direction alternates from site to site. Within the mapped tight-binding framework, the resulting effective Hamiltonian exhibits an SSH-like structure with geometry-induced alternating couplings. Surprisingly, although a spectral gap opens, no localized edge states emerge in the uniform system. We show that this behavior originates from modified boundary parameters generated by the phononic mapping. Introducing a geometric domain wall restores localized interface states inside the gap, which give rise to resonant transmission channels across an otherwise insulating frequency window.

These findings show that the tight-binding and graph-based representation of phonons provides a transparent framework together with practical advantages for the study of vibrational transport in complex systems. Beyond reproducing the original lattice dynamics, the approach reveals geometry-induced effective Hamiltonians, localized domain-wall states, and their associated transport resonances. By bridging methods from electronic transport and lattice dynamics, this framework offers a route toward the analysis and design of phononic structures with controlled transport and localization properties.

\section{Representations of the Tight-Binding Hamiltonian}

The tight-binding (TB) formalism provides a general framework to describe linear wave-like excitations in discrete systems \cite{Datta,Klimeck}. Although most commonly used in the context of electronic structure, its mathematical formulation applies broadly to any system governed by linear coupling between localized degrees of freedom. In this section, we present three equivalent representations of the tight-binding Hamiltonian—operator (ket--bra), matrix, and graph—and emphasize how each highlights different structural aspects of the problem. This unified perspective will serve as the foundation for mapping vibrational systems onto tight-binding models.

\subsection{Equivalent representations: operator, matrix, and graph}

We begin with the operator formulation. Consider a set of localized states $\{ |i\rangle \}$ forming a basis of a finite-dimensional Hilbert space. The tight-binding Hamiltonian can be written as
\begin{equation}
\hat{H} = \sum_i \varepsilon_i |i\rangle \langle i| + \sum_{i \neq j} t_{ij} |i\rangle \langle j| ,
\end{equation}
where $\varepsilon_i$ denotes the onsite energy associated with state $|i\rangle$, and $t_{ij}$ is the hopping amplitude between states $|j\rangle$ and $|i\rangle$. Hermiticity requires $t_{ij} = t_{ji}^*$. This representation makes explicit the operator nature of the problem and is particularly well suited for discussing symmetries, basis transformations, and spectral properties. Importantly, the index $i$ labeling the basis states does not need to correspond strictly to spatial sites, but more generally to a set of localized degrees of freedom.

Upon choosing the basis $\{ |i\rangle \}$, the Hamiltonian is represented by a matrix $\mathbf{H}$ with elements $H_{ij} = \langle i | H | j \rangle$, given by
\begin{equation}
H_{ij} =
\begin{cases}
\varepsilon_i, & i = j, \\
t_{ij}, & i \neq j.
\end{cases}
\end{equation}
The eigenvalue problem
\begin{equation}\label{EqMatrix}
\mathbf{H}\boldsymbol{\psi} = E \boldsymbol{\psi}
\end{equation}
determines the allowed modes of the system, where $\psi_i = \langle i | \psi \rangle$ are the amplitudes in the chosen basis. From this viewpoint, the tight-binding Hamiltonian is a structured matrix whose nonzero elements encode the connectivity between degrees of freedom. This representation is particularly suitable for numerical computations and establishes a direct connection with standard linear algebra techniques.

An equivalent and often more intuitive representation is obtained by interpreting the Hamiltonian as a weighted graph. Each basis state $|i\rangle$ is associated with a node, while each nonzero hopping amplitude $t_{ij}$ defines an edge connecting nodes $i$ and $j$ with weight $t_{ij}$. Onsite energies $\varepsilon_i$ are naturally associated with node attributes. In this picture, the tight-binding Hamiltonian can be viewed as a weighted adjacency operator acting on the graph. The system’s spectral and transport properties are thus directly linked to the graph structure, making this representation particularly useful for visualizing connectivity, identifying propagation pathways, and analyzing systems with nontrivial topology or geometry.

\subsection{Multiple degrees of freedom and block structure}

\begin{figure}[tb]
    \centering
    \includegraphics[scale=0.6]{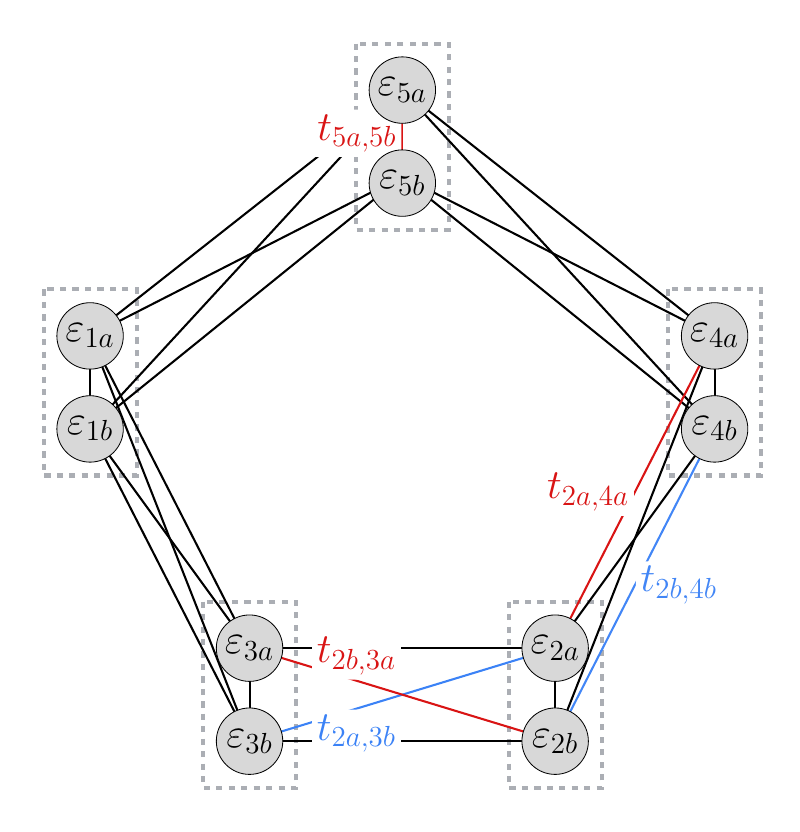} 
    \caption{Graph representation of a multi-orbital tight-binding Hamiltonian defined on a pentagonal lattice. Each site is expanded into two internal degrees of freedom ($a$ and $b$), represented by individual nodes with onsite energies $\varepsilon_{i\alpha}$. In this example, hopping amplitudes are nonzero only between nearest-neighbor sites and include both intra- and inter-orbital couplings. Only a subset of hopping terms is explicitly labeled for clarity.}
    \label{fig:graph}
\end{figure}

Many physical systems contain multiple degrees of freedom per spatial site. Within the tight-binding framework, this situation is naturally described by introducing a set of basis states $\{ |i,\alpha\rangle \}$, where $i$ labels the site and $\alpha$ labels internal degrees of freedom. The Hamiltonian then takes the form
\begin{equation}
H = \sum_{i,\alpha} \varepsilon_{i\alpha} |i,\alpha\rangle \langle i,\alpha|
+ \sum_{i \neq j} \sum_{\alpha,\beta} t_{i\alpha,j\beta} |i,\alpha\rangle \langle j,\beta| ,
\end{equation}
which corresponds, in matrix form, to a block-structured operator with intra-site and inter-site couplings. The corresponding graph representation, making this structure explicit in terms of connectivity between internal degrees of freedom, is shown in Fig.~\ref{fig:graph}.

This multi-orbital structure provides a natural description of systems in which internal degrees of freedom are coupled both locally and between neighboring sites. As we will show below, vibrational systems fall precisely into this category.

\subsection{Connection to vibrational systems}

The equivalence between the representations discussed above is purely algebraic, but it carries direct physical implications. Any system governed by a linear set of coupled equations can, in principle, be recast as a tight-binding model on an appropriate graph. In particular, mass-weighted force constant matrix governing harmonic lattice vibrations defines a linear operator acting on the space of displacements, with a structure formally analogous to a tight-binding Hamiltonian.

This observation provides a natural route to reinterpret vibrational systems within the tight-binding framework, where the identification of degrees of freedom, couplings, and connectivity becomes explicit. In this context, the graph representation offers a particularly transparent visualization of how geometry and interactions shape the structure of the underlying operator.

In the following section, we exploit this correspondence to construct an explicit mapping between phononic systems and multi-orbital tight-binding models. Within this mapping, the cartesian components of the displacement field will play the role of internal orbitals, while elastic couplings give rise to structured hopping matrices determined by the geometry of the system.

\section{Mass-weighted force constant matrix as a multi-orbital tight-binding Hamiltonian}

In this section, we establish an explicit correspondence between lattice vibrational systems and tight-binding models with multiple internal degrees of freedom. This mapping allows us to reinterpret the mass-weighted force constant matrix governing harmonic motion as an effective tight-binding Hamiltonian defined on a graph, where geometry and elastic interactions determine the structure of the effective couplings.

\subsection{Equations of motion and mass-weighted force constant matrix}

We consider a system of point masses connected by elastic interactions within the harmonic approximation. Let $\mathbf{u}_i$ denote the displacement vector of the $i$-th mass from its equilibrium position. The equations of motion are given by
\begin{equation}
m_i \ddot{\mathbf{u}}_i(t) = - \sum_j \mathbf{K}_{ij} \mathbf{u}_j(t) ,
\end{equation}
where $\mathbf{K}_{ij}$ are force-constant matrices encoding the elastic coupling between masses $i$ and $j$.

Assuming time-harmonic solutions of the form $\mathbf{u}_i(t) = \mathbf{u}_i e^{i\omega t}$, and introducing mass-weighted coordinates $\mathbf{v}_i = \sqrt{m_i}\,\mathbf{u}_i$, the equations reduce to

\begin{equation}
\sum_j \mathbf{\Phi}_{ij} \mathbf{v}_j = \omega^2 \mathbf{v}_i,
\end{equation}
where $\mathbf{\Phi}_{ij} = \mathbf{K}_{ij}/\sqrt{m_i m_j}$ defines the mass-weighted force-constant matrix.

Collecting all degrees of freedom into a single vector $\mathbf{v}$, the problem can be written compactly as
\begin{equation}
\mathbf{\Phi}\, \mathbf{v} = \omega^2 \mathbf{v}.
\end{equation}
This equation has the same structure as the tight-binding eigenvalue problem of Eq.~\ref{EqMatrix}, allowing $\mathbf{\Phi}$ to be interpreted as an effective Hamiltonian and $\omega^2$ as the corresponding eigenvalue.

\subsection{Mapping to a tight-binding Hamiltonian}

The formal analogy between $\mathbf{\Phi}$ and a tight-binding Hamiltonian can be made explicit by identifying the components of $\mathbf{v}$ as basis amplitudes. We introduce a set of states $\{ |i,\alpha\rangle \}$, where $i$ labels the mass and $\alpha \in \{x,y,z\}$ the Cartesian components of its displacement.

In this basis, the mass-weighted force-constant matrix defines the operator
\begin{equation}
\hat{\Phi} = \sum_{i,\alpha} \varepsilon_{i\alpha} |i,\alpha\rangle \langle i,\alpha|
+ \sum_{i,j} \sum_{\alpha,\beta} t_{i\alpha,j\beta} |i,\alpha\rangle \langle j,\beta|,
\end{equation}
where $t_{i\alpha,i\alpha}=0$. Off-diagonal terms describe both intra-site ($i=j$, $\alpha \ne \beta$) and inter-site ($i \ne j$) couplings. The onsite terms $\varepsilon_{i\alpha}$ and hopping amplitudes $t_{i\alpha,j\beta}$ are determined by the elements of the force-constant matrices $\mathbf{K}_{ij}$, written in the composite index form $i\alpha$:
\begin{equation}
\varepsilon_{i\alpha} = \frac{K_{i\alpha,i\alpha}}{m_i}, 
\qquad
t_{i\alpha,j\beta} = \frac{K_{i\alpha,j\beta}}{\sqrt{m_i m_j}}.
\label{Eq:Hopping}
\end{equation}

Within this representation, elastic interactions generate hopping terms both between sites and between internal degrees of freedom. In particular, off-diagonal elements in the Cartesian indices ($\alpha \neq \beta$) describe coupling between different Cartesian components, which do not necessarily coincide with the physical polarization directions defined by the geometry of the system. Such couplings reflect mixing at the level of the chosen representation and do not, in general, imply mixing of the underlying longitudinal and transverse modes.

This mapping naturally defines a graph representation in which each degree of freedom $(i,\alpha)$ corresponds to a node and each nonzero element $\Phi_{i\alpha,j\beta}$ defines a weighted edge. Elastic interactions thus generate a structured connectivity between internal degrees of freedom, reflecting both the geometry of the system and the pathways for vibrational propagation as encoded in the chosen representation. 

This representation depends on the choice of cartesian axes and transforms under rotations of the coordinate system, although the underlying physical properties remain invariant.

\subsection{Two-mass building block}

To illustrate this construction, we consider a minimal building block consisting of two masses coupled by anisotropic elastic interactions. This unit serves as the basis for the more complex systems studied later, which can be constructed as assemblies of such elements and naturally described in terms of graph connectivity.

We first introduce a local reference frame in which the $x$-axis is aligned with the direction connecting the two masses. The remaining axes are chosen along the principal directions of the transverse response, so that the interaction matrix is diagonal in this basis. This choice naturally extends to both two- and three-dimensional systems, where the longitudinal direction is defined by the bond and the remaining directions span the transverse subspace. In this frame, the longitudinal force constant is denoted by $\alpha$, while $\beta$ and $\gamma$ correspond to transverse force constants. The interaction can then be written as
\begin{equation}
    \mathbf{V} =
    \begin{pmatrix}
        \alpha & 0 & 0 \\
        0 & \beta & 0 \\
        0 & 0 & \gamma
    \end{pmatrix}.
\end{equation}

The matrix $\mathbf{V}$ represents the local force-constant matrix in the principal frame of the interaction. To describe a general configuration, we introduce a rotation $\mathbf{R}$ that relates this local frame to the global Cartesian frame, leading to the transformed interaction matrix
\begin{equation}
    \mathbf{V}' = \mathbf{R} \mathbf{V} \mathbf{R}^T.
\end{equation}

The matrix $\mathbf{V}'$ determines the corresponding force-constant blocks for the pair of masses,
\begin{equation}
\mathbf{K}_{12} = -\mathbf{V}', \qquad \mathbf{K}_{21} = -\mathbf{V}',
\end{equation}
while the onsite blocks follow from the structure of the force-constant matrix as
\begin{equation}
\mathbf{K}_{11} = \mathbf{V}', \qquad \mathbf{K}_{22} = \mathbf{V}'.
\end{equation}

By subsituting in Eq.~\ref{Eq:Hopping}, these blocks define the hopping amplitudes $t_{i\alpha,j\beta}$ of the effective tight-binding model. The elements of $\mathbf{V}'$ therefore map directly onto couplings between internal degrees of freedom: diagonal terms couple identical displacement components, while off-diagonal terms ($\alpha \neq \beta$) generate inter-orbital mixing.

To make this structure explicit, we consider a rotation by an angle $\theta$ in the plane defined by the longitudinal direction and one transverse direction. Restricting to this two-dimensional subspace, the interaction matrix in its principal frame reads
\begin{equation}
\mathbf{V} =
\begin{pmatrix}
\alpha & 0 \\
0 & \beta
\end{pmatrix},
\end{equation}
while the rotation is given by
\begin{equation}
\mathbf{R}(\theta) =
\begin{pmatrix}
\cos\theta & -\sin\theta \\
\sin\theta & \cos\theta
\end{pmatrix}.
\end{equation}

The rotated interaction matrix $\mathbf{V}' = \mathbf{R}(\theta)\mathbf{V}\mathbf{R}^T(\theta)$ becomes
\begin{equation}
\mathbf{V}' =
\begin{pmatrix}
\alpha \cos^2\theta + \beta \sin^2\theta & (\alpha - \beta)\sin\theta\cos\theta \\
(\alpha - \beta)\sin\theta\cos\theta & \alpha \sin^2\theta + \beta \cos^2\theta
\end{pmatrix}.
\end{equation}

This result shows that off-diagonal terms arise whenever the interaction is anisotropic ($\alpha \neq \beta$) and the bond is not aligned with the coordinate axes ($\theta \neq 0,\pi/2$). Through the mapping to the tight-binding model, these terms generate hopping amplitudes $t_{i\alpha,j\beta}$ with $\alpha \neq \beta$, reflecting geometry-induced coupling between different displacement components.

\begin{figure*}[h]
\centering
    a)\includegraphics[width=\textwidth]{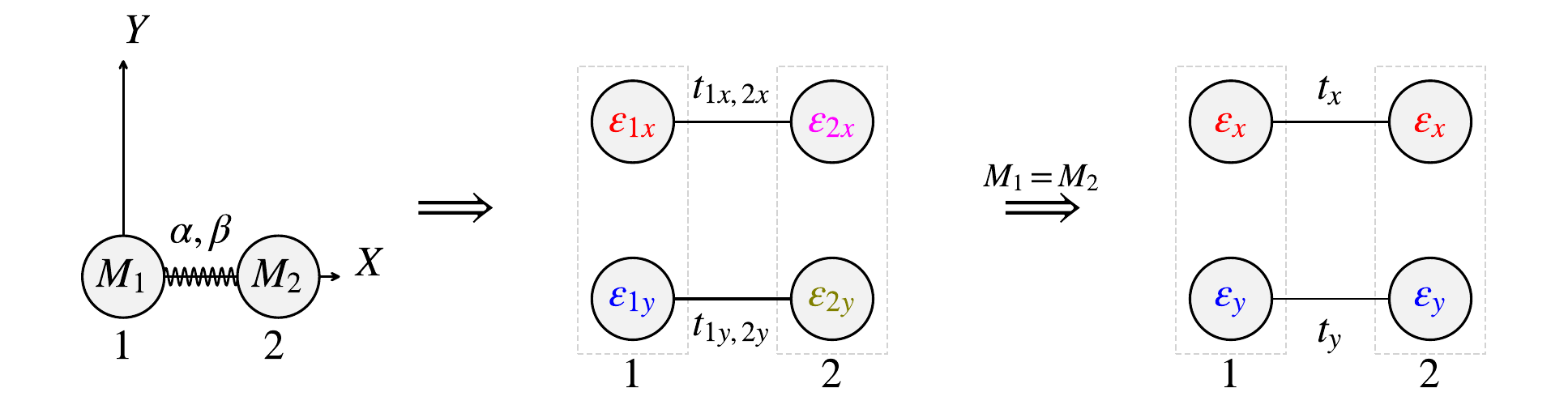}
    b)\includegraphics[width=\textwidth]{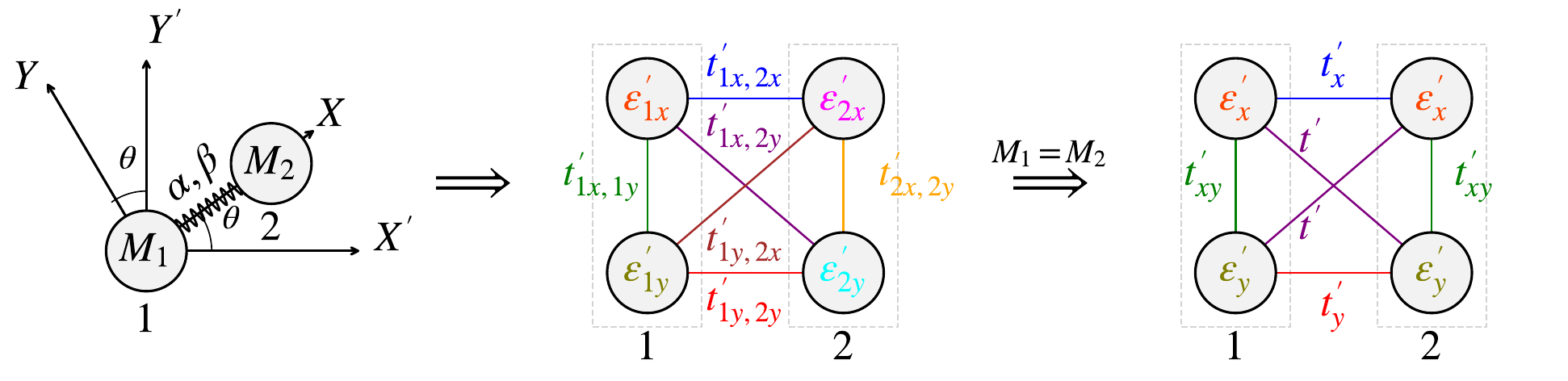}
    \caption{Equivalent tight-binding representation for (a) two coupled masses aligned with the coordinate axes and (b) the same system rotated by an angle $\theta$, illustrating the emergence of inter-orbital couplings.}
    \label{fig:Mapped}
\end{figure*}

The effect of geometric orientation is illustrated in Fig.~\ref{fig:Mapped}. When the interaction is aligned with the Cartesian axes [Fig.~\ref{fig:Mapped}(a)], the interaction matrix remains diagonal and the corresponding tight-binding representation involves couplings only between identical displacement components, without mixing between different degrees of freedom.

In contrast, when the bond is rotated with respect to the coordinate axes [Fig.~\ref{fig:Mapped}(b)], the interaction matrix acquires off-diagonal elements. In the tight-binding representation, these terms generate inter-orbital hopping amplitudes $t_{i\alpha,j\beta}$ with $\alpha \neq \beta$, leading to a fully connected structure between the internal degrees of freedom of neighboring sites.

This behavior reflects the fact that, for anisotropic interactions, the principal directions of the elastic response do not generally coincide with the global Cartesian frame. While this mixing arises from the chosen representation at the level of a single bond, it becomes unavoidable in extended systems where different interactions are associated with incompatible local frames, so that no global basis can simultaneously diagonalize all couplings, leading to effective mixing between physical modes.

\subsection{From building blocks to networks}

The previous construction extends naturally to systems with multiple elastic interactions. Since the equations of motion are linear, the total force-constant matrix is obtained as the superposition of the contributions associated with each individual spring.

\begin{figure*}[h]
\centering
    \includegraphics[width=\textwidth]{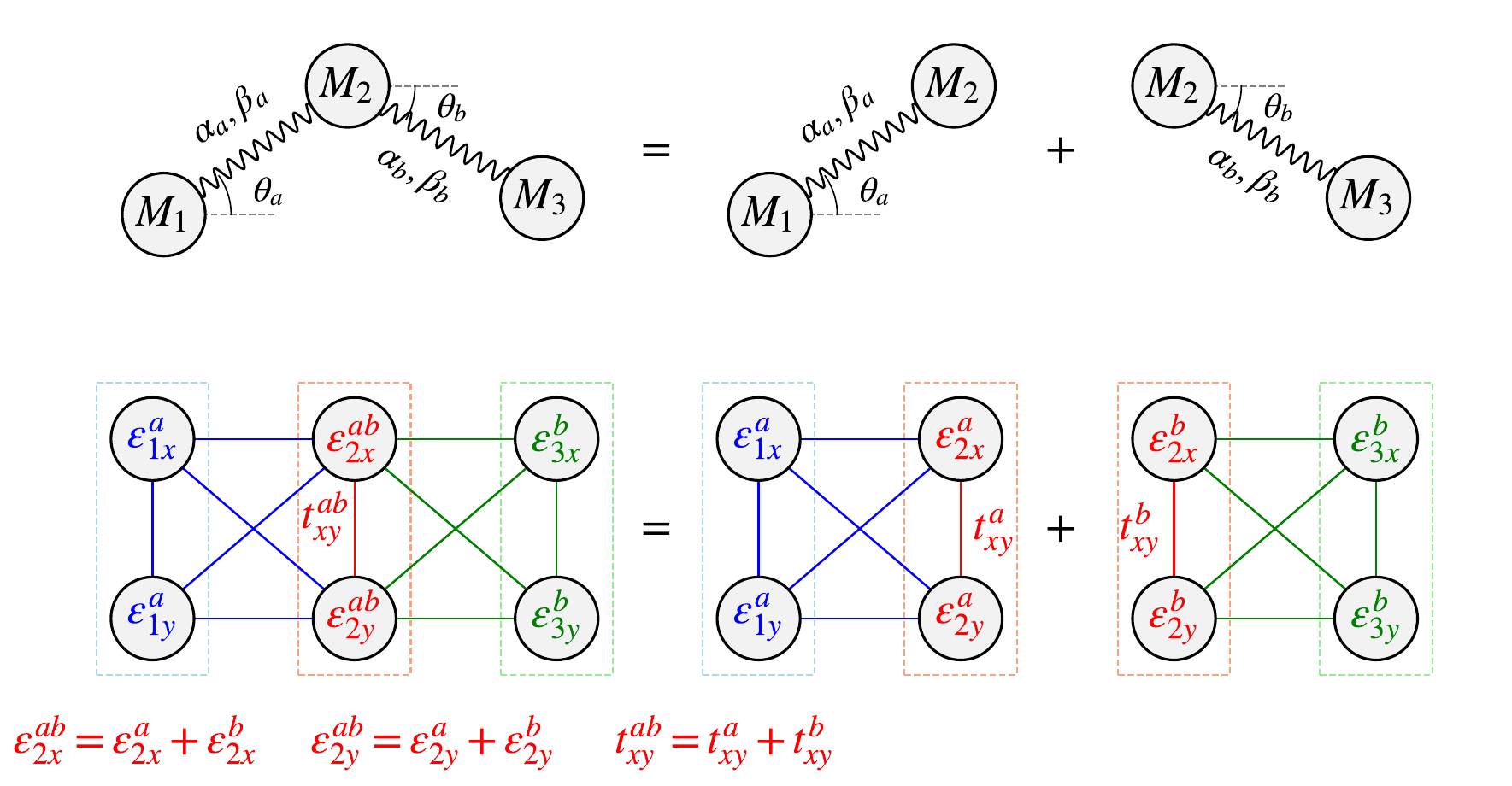}
    \caption{Schematic representation of a three-atom system with different atomic masses connected by different force constants, together with its building-block decomposition and the corresponding mapped tight-binding graph.}
    \label{fig:blocks}
\end{figure*}

In the graph representation, this corresponds to combining the graphs associated with each interaction. Each spring defines a local contribution to the force-constant matrix, and therefore to the effective tight-binding model, acting only on the degrees of freedom of the masses it connects. When several springs are present, the total graph is obtained by assembling these local contributions.

As a concrete example, consider three masses $m_1$, $m_2$, and $m_3$, where $m_1$ and $m_2$ are connected by a spring oriented at a given angle, and $m_2$ and $m_3$ are connected by a second spring with a different orientation. In the two-dimensional case, each mass contributes two internal degrees of freedom, resulting in a graph with six nodes.

The nodes associated with the central mass $m_2$ receive contributions from both interactions. As a result, their onsite energies include the sum of the corresponding terms arising from each spring. In addition, couplings between degrees of freedom belonging to $m_2$, namely intracell hopping terms, also accumulate contributions from the different interactions.

Each interaction contributes independently to the hopping amplitudes connecting the degrees of freedom of the masses it links. Additivity therefore occurs only at the level of matrix elements associated with shared degrees of freedom, while couplings between distinct pairs of masses remain independent. The resulting tight-binding model is thus obtained by superposing these local contributions.

This example illustrates a general principle: complex systems can be constructed by combining elementary graph contributions associated with individual interactions, with onsite terms and hopping amplitudes combining additively whenever they act on the same degrees of freedom.

\section{Linear monoatomic chain with an angular bend}

\begin{figure*}[h]
    \centering
    \includegraphics[width=\textwidth]{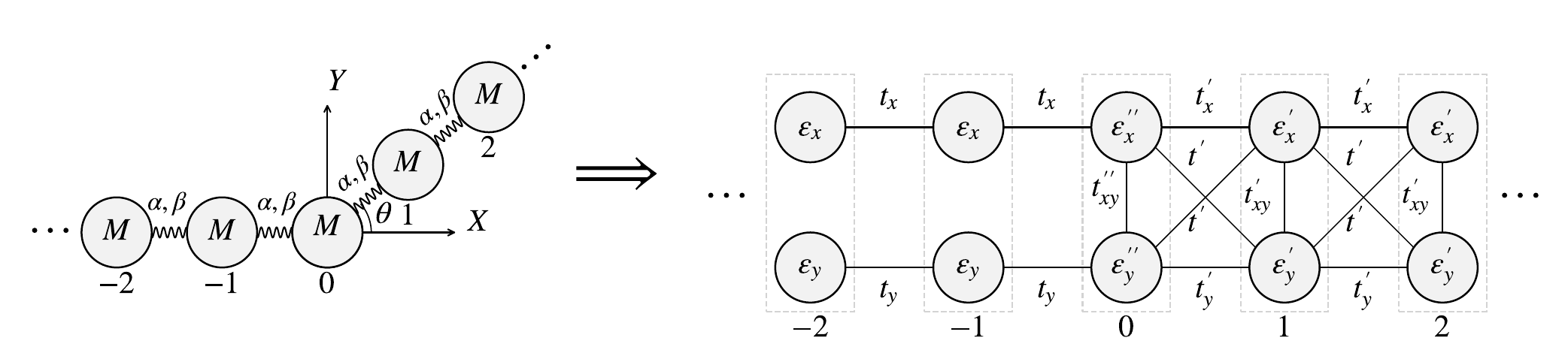}
    \caption{Monoatomic chain with an angular bend and its corresponding mapped multi-orbital tight-binding representation.}
    \label{fig:chain angular bend}
\end{figure*}

To validate the proposed mapping, we consider a monoatomic chain containing a localized angular defect. The system consists of two semi-infinite chains connected at a vertex, where the second branch is rotated by an angle $\theta$ with respect to the first, as shown in Fig.~\ref{fig:chain angular bend}.

This geometry provides an ideal benchmark for the present framework. On the one hand, it is sufficiently simple to admit an analytical solution within conventional lattice dynamics. On the other hand, the angular mismatch induces coupling between vibrational polarizations, producing mode conversion and nontrivial scattering that constitute a stringent test of the mapped tight-binding description.

We restrict the analysis to in-plane vibrations within the $XY$ plane and model the elastic interactions using the nearest-neighbor Born potential

\begin{equation}
\begin{aligned}
V_{j,j+1}
&=
\frac{\alpha-\beta}{2}
\left[
(\mathbf{u}_j-\mathbf{u}_{j+1})
\cdot
\mathbf{n}_{j,j+1}
\right]^2 \\
&+
\frac{\beta}{2}
\left\|
\mathbf{u}_j-\mathbf{u}_{j+1}
\right\|^2,
\end{aligned}
\end{equation}
where $\alpha$ and $\beta$ denote the longitudinal and transverse force constants, respectively \cite{Gonzalez_2022}.

Away from the defect, the system reduces to a uniform one-dimensional chain supporting longitudinal and transverse vibrations with dispersions

\begin{equation}
M\omega^2=
2\alpha(1-\cos k^La),
\end{equation}
and

\begin{equation}
M\omega^2=
2\beta(1-\cos k^Ta),
\end{equation}
respectively. The angular bend acts as a localized scattering center that couples these two polarizations.

\subsection{Analytical lattice-dynamics solution}

The scattering problem is formulated by matching longitudinal and transverse vibrational modes at the defect site, leading to a scattering matrix that relates incoming and outgoing amplitudes. The complete derivation of the scattering amplitudes, including both the propagating and evanescent regimes, is presented in Appendix \ref{AppendixA}.

The scattering amplitudes define transmission and reflection probabilities between the different vibrational polarizations,

\begin{equation}
\begin{aligned}
&T_{L,L}=|S_{LL}^t|^2,\\
&T_{T,L}=\frac{v_T}{v_L}|S_{TL}^t|^2,\\
&T_{L,T}=\frac{v_L}{v_T}|S_{LT}^t|^2,\\
&T_{T,T}=|S_{TT}^t|^2,
\end{aligned}
\end{equation}
and

\begin{equation}
\begin{aligned}
&R_{L,L}=|S_{LL}^r|^2,\\
&R_{T,L}=\frac{v_T}{v_L}|S_{TL}^r|^2,\\
&R_{L,T}=\frac{v_L}{v_T}|S_{LT}^r|^2,\\
&R_{T,T}=|S_{TT}^r|^2.
\end{aligned}
\end{equation}

The first index labels the outgoing polarization, while the second index labels the incoming polarization. In these expressions,

\begin{equation}
v_L=\frac{\partial\omega}{\partial k^L},
\qquad
v_T=\frac{\partial\omega}{\partial k^T},
\end{equation}
are the longitudinal and transverse group velocities, respectively, given by

\begin{equation}
v_L=
a\sqrt{\frac{\alpha}{M}-\frac{\omega^2}{4}},
\qquad
v_T=
a\sqrt{\frac{\beta}{M}-\frac{\omega^2}{4}}.
\end{equation}

The velocity ratios appear because the scattering matrix relates wave amplitudes, whereas transmission and reflection probabilities must be defined in terms of the associated flux carried by each propagating mode.

Flux conservation requires

\begin{equation}
T_{L,L}+T_{T,L}+R_{L,L}+R_{T,L}=1,
\end{equation}
for a longitudinally incident wave, and

\begin{equation}
T_{L,T}+T_{T,T}+R_{L,T}+R_{T,T}=1,
\end{equation}
for a transversely incident wave.

\begin{figure*}[tb]
    \centering
    \includegraphics[width=\textwidth]{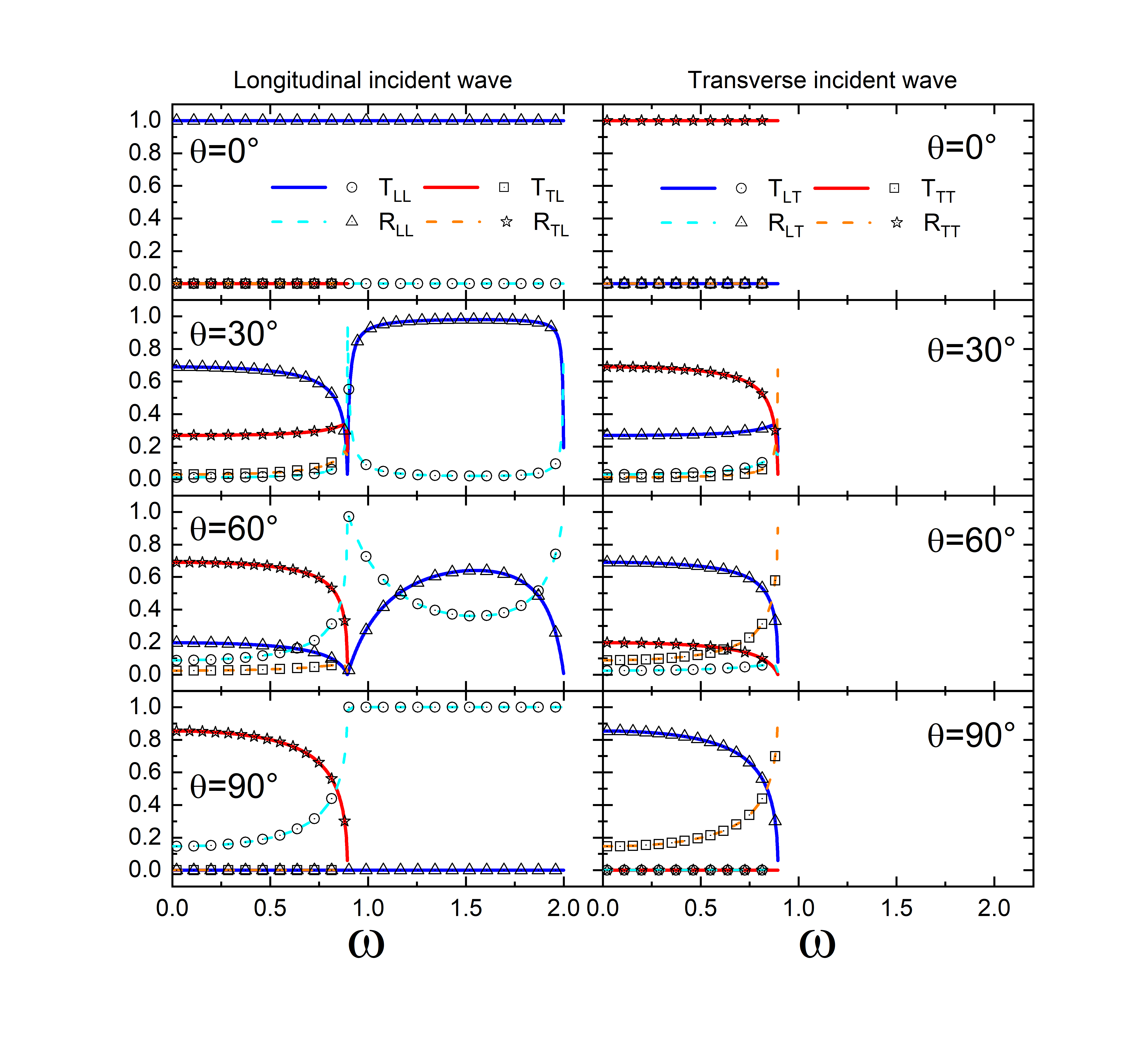}
    \caption{Transmission and reflection probabilities for a monoatomic chain with an angular bend as a function of frequency. Solid lines correspond to the analytical lattice-dynamics solution. Symbols indicate the results obtained from the recursive scattering-matrix calculation performed on the mapped tight-binding model.}
    \label{fig:TR}
\end{figure*}

The analytical solution provides direct access to the polarization-resolved scattering coefficients and therefore allows the role of geometry-induced mode conversion to be analyzed in detail. Figure~\ref{fig:TR} presents the resulting transmission and reflection probabilities as functions of frequency for several bending angles, namely $\theta=0^\circ$, $30^\circ$, $60^\circ$, and $90^\circ$, considering the representative case $\beta=0.2\alpha$. The left column of Fig.~\ref{fig:TR} corresponds to longitudinal incidence, whereas the right column corresponds to transverse incidence.

For $\theta=0^\circ$, the chain remains perfectly straight and no scattering takes place. Consequently, transmission is perfect within the incident polarization channel and reflection vanishes identically. Longitudinal incidence yields $T_{L,L}=1$, while transverse incidence yields $T_{T,T}=1$. All remaining transmission and reflection probabilities are zero.

For intermediate angles, such as $\theta=30^\circ$ and $\theta=60^\circ$, the scattering process distributes the energy among multiple outgoing channels. The relative magnitude of the polarization-resolved probabilities changes continuously with both frequency and bending angle, indicating that the angular defect acts as a frequency-dependent mode converter. In particular, the conversion efficiency increases as the bending angle increases, leading to a stronger redistribution of spectral weight between longitudinal and transverse channels.

A comparison between longitudinal and transverse incidence reveals an exact correspondence between the polarization-resolved scattering coefficients within the frequency range where both modes are propagating. In particular,

\begin{equation}
    T_{L,L}=T_{T,T},
    \qquad
    T_{T,L}=T_{L,T},
\end{equation}
and similarly

\begin{equation}
  R_{L,L}=R_{T,T},
    \qquad
    R_{T,L}=R_{L,T}.  
\end{equation}

The transmission equalities can be derived directly from the analytical expressions of the scattering amplitudes and are demonstrated in Appendix \ref{AppendixB}. These identities hold over the entire frequency range where both polarization channels remain propagating. Consequently,

\begin{equation}
    T_{L,L}+T_{T,L}
    =
    T_{L,T}+T_{T,T},
\end{equation}
and

\begin{equation}
    R_{L,L}+R_{T,L}
    =
    R_{L,T}+R_{T,T},
\end{equation}
so that the total transmission and reflection probabilities are identical for longitudinal and transverse incidence whenever both polarization channels remain propagating. This symmetry reflects the equivalent role played by the longitudinal and transverse channels in the mode-conversion process induced by the angular defect.

The strongest polarization conversion occurs for $\theta=90^\circ$, where the two branches become orthogonal. In this limit, the defect acts as a perfect polarization converter. For longitudinal incidence, all propagating transmission occurs through the transverse channel while reflection remains purely longitudinal, implying $T_{L,L}=0$ and $R_{T,L}=0$. Conversely, for transverse incidence, transmission occurs exclusively through the longitudinal channel while reflection remains purely transverse, yielding $T_{T,T}=0$ and $R_{L,T}=0$. This behavior highlights the purely geometric origin of the mode-conversion mechanism.

Another important feature emerges at the transverse band edge,

\begin{equation}
\omega_{\mathrm{max},T}
=
2\sqrt{\beta/M}
=
2\sqrt{0.2}\,\omega_0,
\end{equation}
where $\omega_0=\sqrt{\alpha/M}$. The frequency range of the transverse-incidence calculations is therefore restricted to $\omega\leq\omega_{\mathrm{max},T}$, since no propagating transverse incident states exist above the transverse band edge.

For frequencies above $\omega_{\mathrm{max},T}$, transverse propagating modes cease to exist and the transverse channel becomes evanescent. Consequently, all transmission and reflection probabilities associated with propagating transverse output channels vanish.

Nevertheless, the evanescent transverse modes remain coupled to the junction and therefore continue to influence the scattering process. As a result, the longitudinal transmission and reflection probabilities retain a nontrivial dependence on the transverse degree of freedom even when no transverse propagating states exist.

For longitudinal incidence, the probabilities $T_{L,L}$ and $R_{L,L}$ continue to be modified by the coupling to the localized evanescent transverse mode even after the transverse channel ceases to propagate. Likewise, for transverse incidence, all propagating transmission channels available to a transversely incident wave collapse at the transverse band edge, leading to complete reflection. As the frequency approaches $\omega_{\mathrm{max},T}$, transmission is progressively suppressed while reflection approaches unity for all nonzero bending angles. This behavior illustrates that evanescent channels may play a decisive role in phonon transport even when they no longer contribute directly to propagation.

\subsection{Mapped tight-binding description and recursive scattering matrix method}

The same system can be described within the mapped multi-orbital tight-binding representation shown in Fig.~\ref{fig:chain angular bend}. Under this mapping, the angular bend is transformed into a localized scattering region connecting two semi-infinite periodic leads.

The left lead is diagonal in the orbital basis and consists of two independent channels associated with the Cartesian displacement components. The rotated branch, in contrast, contains geometry-induced inter-orbital couplings generated by the local rotation of the interaction matrices. Consequently, the original phonon scattering problem is recast as a multichannel transport problem formally equivalent to those encountered in electronic tight-binding systems.

The parameters of the effective Hamiltonian follow directly from the construction developed in Sec.~III. In particular, the onsite energies and hopping amplitudes are obtained through the superposition rules associated with the building-block decomposition. Within this representation, the angular defect appears simply as a localized modification of the graph connectivity and hopping structure.

Once the effective Hamiltonian has been constructed, transport properties can be calculated using standard techniques originally developed for electronic systems. Here, we employ the recursive scattering-matrix method \cite{Rodriguez} to determine the scattering amplitudes of the mapped graph.

The recursive procedure first determines the propagating eigenmodes of the semi-infinite leads and subsequently computes the scattering probabilities between them \cite{Ramirez}. These transport channels emerge naturally from the effective Hamiltonian and are therefore identified by their mode index rather than by a priori longitudinal or transverse labels. Nevertheless, for the benchmark system considered here, the propagating channels of the effective leads coincide with the longitudinal and transverse polarizations of the original vibrational problem. This allows a direct comparison between the polarization-resolved scattering coefficients obtained analytically and those computed using the recursive scattering-matrix method.

The symbols shown in Fig.~\ref{fig:TR} correspond to the recursive scattering-matrix calculation performed on the mapped tight-binding model, whereas the continuous lines correspond to the analytical lattice-dynamics solution. The recursive calculation reproduces the analytical results to numerical precision, with discrepancies below $10^{-10}$ for all bending angles and frequencies considered.

Importantly, this agreement is not restricted to the total transmission or reflection probability. The recursive calculation reproduces individually all polarization-resolved transmission and reflection probabilities, including both polarization-preserving processes ($T_{L,L}$, $T_{T,T}$, $R_{L,L}$, and $R_{T,T}$) and polarization-conversion processes ($T_{T,L}$, $T_{L,T}$, $R_{T,L}$, and $R_{L,T}$). The agreement persists in the vicinity of the transverse band edge and throughout the regime where evanescent transverse modes become relevant.

The exact overlap between the numerical results and the analytical curves demonstrates that the mapped tight-binding representation faithfully reproduces the scattering properties of the original phononic system. In particular, the effective model correctly captures mode conversion, angular dependence, conservation of flux, and the influence of evanescent channels on the scattering process.

More importantly, the validation shown in Fig.~\ref{fig:TR} establishes that phonon transport can be analyzed using the same computational machinery routinely employed in electronic transport calculations once the mapping has been constructed. This includes not only the calculation of total transmission and reflection probabilities, but also the resolution of individual scattering channels. The original vibrational problem is therefore reduced to a transport problem on an effective graph without loss of physical information.

Although the angular bend remains sufficiently simple to admit an analytical treatment, the mapped formulation becomes especially valuable for systems in which such analytical solutions are no longer available. This motivates the application of the method to extended structures with nontrivial connectivity, which we investigate in the following section.

\section{Zigzag chains and geometry-induced topological effects}

The angular bend considered in the previous section provides a stringent validation of the mapped tight-binding description, since the transport properties can be compared directly against an exact analytical solution. Having established the validity of the mapping, we now turn to systems for which analytical treatments become considerably more difficult.

A particularly interesting class of examples is provided by zigzag chains, where the orientation of successive bonds alternates throughout the structure. In such systems, the local principal axes associated with neighboring interactions are different, leading to a systematic mixing between longitudinal and transverse vibrational polarizations. Within the mapped tight-binding representation, this geometric effect generates alternating effective hopping amplitudes and therefore gives rise to Hamiltonians that closely resemble one-dimensional dimerized models.

This observation suggests a connection with the Su--Schrieffer--Heeger (SSH) model \cite{SSH}, where alternating hopping amplitudes produce topological phases characterized by the appearance of localized edge states. An important question is therefore whether analogous states can emerge purely as a consequence of the geometry of a vibrational system.

\begin{figure*}[tb]
    \centering
    \includegraphics[width=\textwidth]{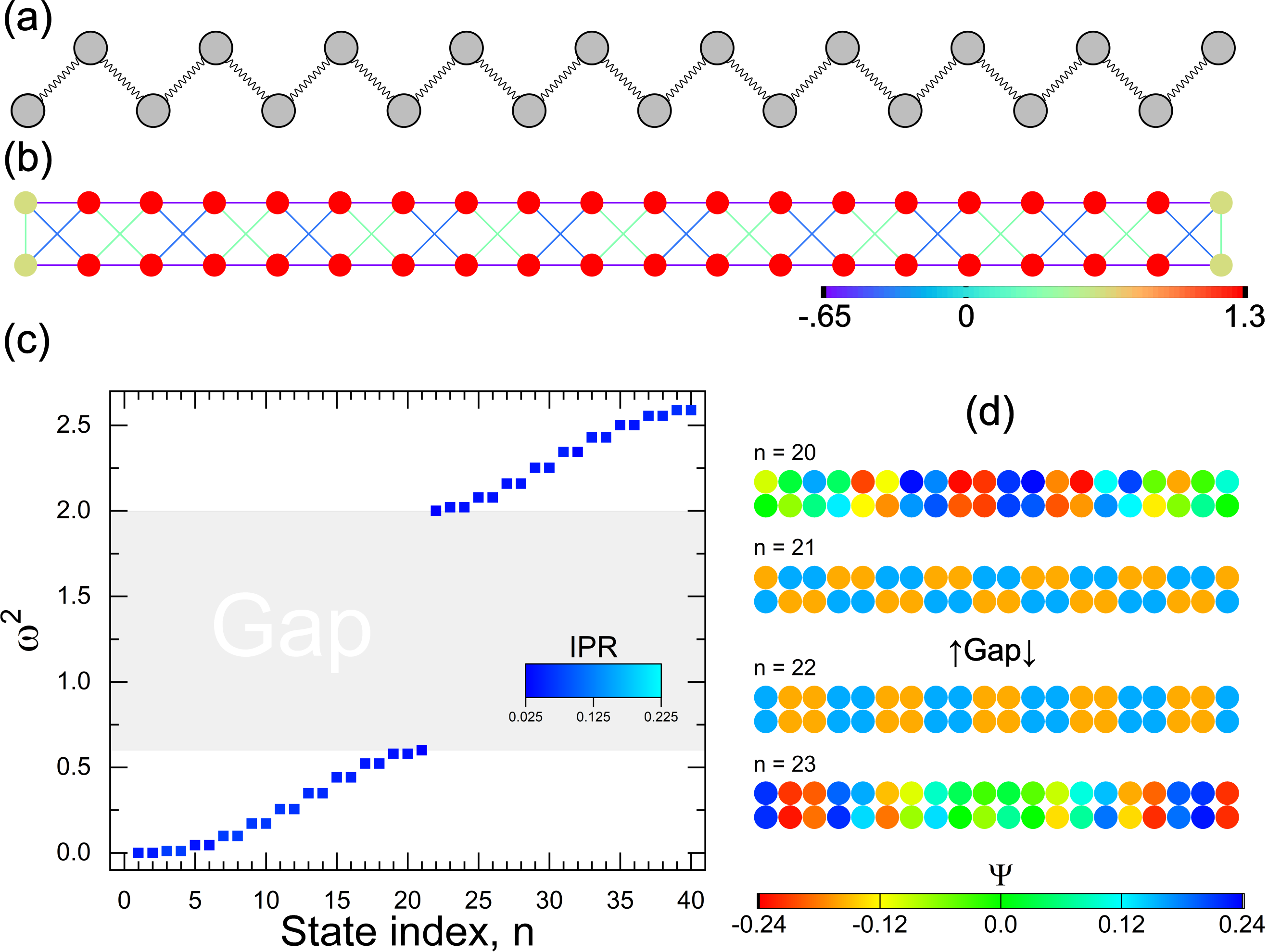}
    \caption{(a) Zigzag monoatomic chain with identical springs. (b) Corresponding mapped multi-orbital tight-binding representation. Colors indicate the values of the effective onsite energies and hopping amplitudes. (c) Spectrum of the effective Hamiltonian as a function of the state index $n$. Colors represent the inverse participation ratio (IPR). A spectral gap opens between the two groups of states. (d) Representative normalized eigenstates. States $n=21$ and $n=22$ delimit the gap and remain spatially extended despite the SSH-like structure of the effective model.}
    \label{fig:zigzag}
\end{figure*}

\subsection{Uniform zigzag chain}

We begin by considering a monoatomic zigzag chain composed of identical elastic interactions, as shown in Fig.~\ref{fig:zigzag}(a). The geometry is characterized by the opening angle between consecutive bonds and by the number of unit cells. Throughout this subsection we consider the representative case $\alpha=1$ and $\beta=0.3$, with neighboring bonds arranged orthogonally.

In this configuration, the longitudinal polarization associated with a given spring becomes coupled to the transverse polarization of the following spring, while the transverse polarization couples to the longitudinal polarization of its neighbor. Consequently, the vibrational dynamics can be effectively viewed as two alternating sequences of spring constants. One of these follows the sequence

\[
\alpha-\beta-\alpha-\beta-\cdots-\alpha ,
\]
whereas the other follows

\[
\beta-\alpha-\beta-\alpha-\cdots-\beta .
\]

At first sight, this structure closely resembles the SSH model. Since $\alpha>\beta$, one would therefore expect the appearance of localized states inside the spectral gap associated with the nontrivial dimerization pattern \cite{Asboth2016}.

The mapped tight-binding Hamiltonian corresponding to the Cartesian frame used in Fig.~\ref{fig:zigzag}(a) can be identified directly in Fig.~\ref{fig:zigzag}(b), where the alternating effective hopping amplitudes give rise to two decoupled dimerized chains, while the end sites exhibit modified onsite energies.

These boundary modifications have a simple physical origin. Each spring contributes simultaneously to the hopping amplitudes between neighboring degrees of freedom and to the onsite terms associated with the masses it connects. In the bulk of the chain, every mass participates in two neighboring interactions and therefore receives contributions from both adjacent springs. At the system boundaries, however, one of these interactions is missing. As a result, the corresponding onsite terms differ from those in the bulk, producing the modified boundary parameters visible in Fig.~\ref{fig:zigzag}(b).

To quantify the degree of spatial localization of the eigenstates, we evaluate the inverse participation ratio (IPR),

\begin{equation}
\mathrm{IPR}
=
\sum_i |\psi_i|^4 ,
\end{equation}
for normalized eigenvectors satisfying
$\sum_i |\psi_i|^2=1$.
Extended states exhibit IPR values of the order of $1/N$, where $N$ is the number of degrees of freedom of the system, whereas localized states exhibit comparatively large IPR values. This quantity therefore provides a convenient measure of localization in the effective model.

Figure~\ref{fig:zigzag}(c) shows the spectrum of the effective Hamiltonian together with the inverse participation ratio (IPR) of each eigenstate. A well-defined spectral gap is observed, confirming that the alternating couplings strongly modify the vibrational spectrum. Surprisingly, however, no eigenstates appear inside the gap. An additional feature of the spectrum is the presence of pairwise degeneracies. All eigenvalues occur in degenerate pairs, with numerical agreement extending to at least twelve significant digits, except for the states $n=21$ and $n=22$ that delimit the spectral gap.

This absence of in-gap states is reinforced by the analysis of the eigenvectors shown in Fig.~\ref{fig:zigzag}(d). The states $n=21$ and $n=22$, which delimit the gap, remain spatially extended over the entire system rather than becoming localized at the boundaries. The IPR provides a quantitative confirmation of this behavior, with all eigenstates exhibiting values of the order of $1/N$, characteristic of extended bulk states, and the states immediately below and above the gap are among the most delocalized states of the spectrum.

Therefore, despite the SSH-like alternation of effective couplings and the appearance of a spectral gap, the mapped zigzag chain does not support edge states. This result indicates that the modified boundary parameters generated by the phononic mapping alter the ideal SSH picture and fully suppress the emergence of localized edge modes.

\begin{figure*}[!tb]
    \centering
    \includegraphics[width=\textwidth]{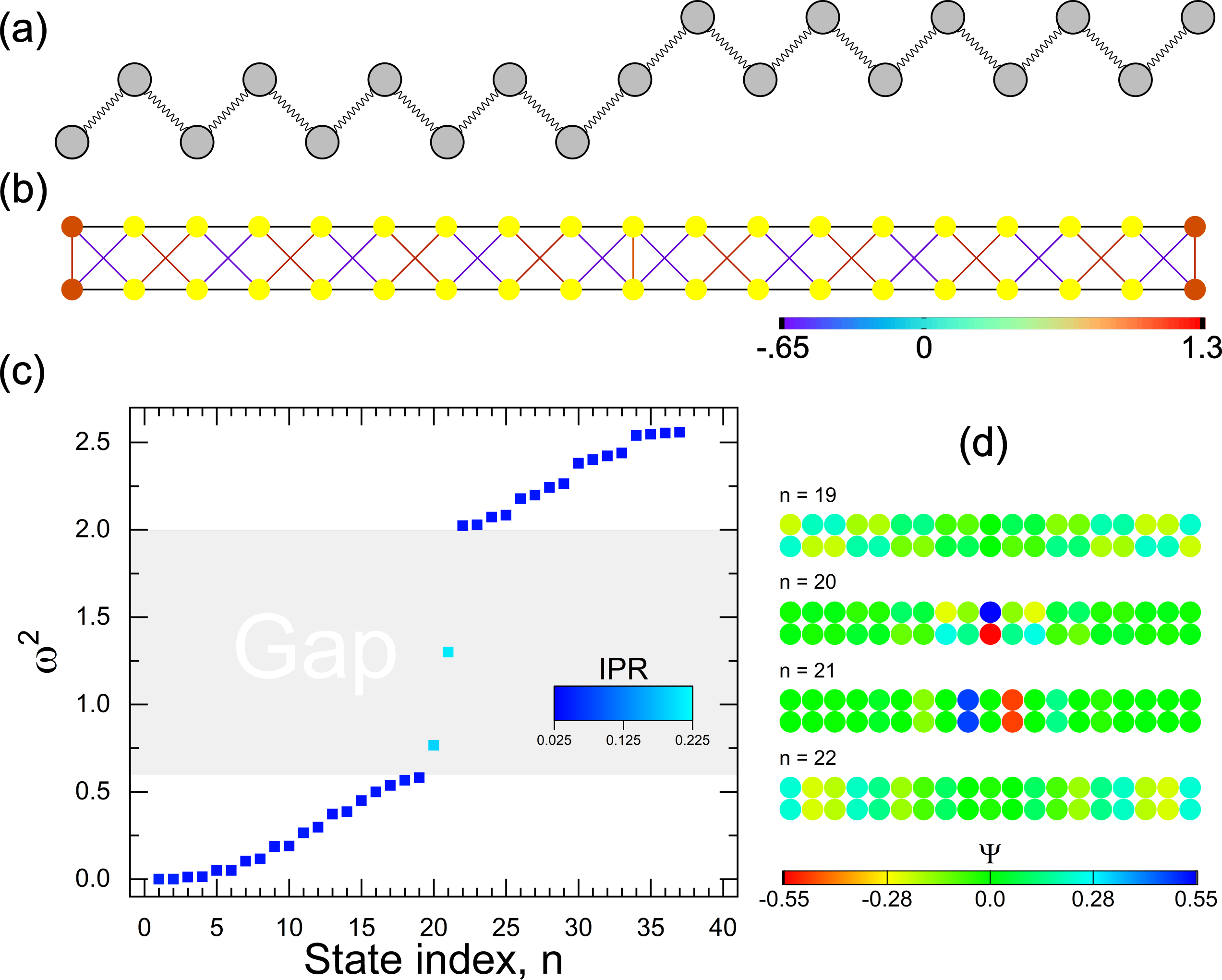}
    \caption{(a) Zigzag chain containing a geometric domain wall. The domain wall is formed by two consecutive bonds with the same orientation, producing a local inversion of the dimerization pattern. (b) Corresponding mapped multi-orbital tight-binding representation. Colors indicate the values of the effective onsite energies and hopping amplitudes. (c) Spectrum of the effective Hamiltonian as a function of the state index $n$. The shaded region corresponds to the same energy interval highlighted in Fig.~\ref{fig:zigzag}(c). Colors represent the inverse participation ratio (IPR). Two localized states emerge inside the spectral gap. (d) Representative normalized eigenstates. States $n=20$ and $n=21$ are localized around the domain wall and appear inside the spectral gap, whereas states $n=19$ and $n=22$ remain extended bulk states.}
    \label{fig:DW}
\end{figure*}

\subsection{Domain-wall zigzag chain}

We now introduce a geometric domain wall by reversing the zigzag pattern at the center of the chain, as shown in Fig.~\ref{fig:DW}(a). The domain wall is formed when two consecutive springs share the same orientation, creating a local interruption of the alternating geometric pattern present in the uniform zigzag chai. Such domain walls are known to play a central role in the emergence of localized states in dimerized tight-binding systems \cite{munoz,Zurita}.

Within the effective description, this modification generates two dimerized sequences of spring constants,

\[
\alpha-\beta-\alpha-\cdots-\alpha
\;\;|\;\;
\alpha-\beta-\cdots-\alpha ,
\]
and

\[
\beta-\alpha-\beta-\cdots-\beta
\;\;|\;\;
\beta-\alpha-\cdots-\beta ,
\]
where the vertical bar indicates the location of the domain wall. In both cases, the alternating pattern is interrupted by the repetition of two consecutive couplings, giving rise to a local inversion of the dimerization pattern.

The corresponding mapped tight-binding Hamiltonian is shown in Fig.~\ref{fig:DW}(b). The domain wall can be identified directly through the modified hopping sequence at the center of the graph, where consecutive effective hoppings are repeated. Unlike the physical boundaries discussed in the previous subsection, the onsite energies remain uniform in the vicinity of the domain wall. The local modification is therefore encoded primarily in the hopping structure of the effective Hamiltonian.

The resulting spectrum is presented in Fig.~\ref{fig:DW}(c). The shaded region corresponds to the same energy interval highlighted in Fig.~\ref{fig:zigzag}(c). The bulk spectrum remains largely unchanged, indicating that the domain wall acts as a local perturbation that preserves the overall band structure of the system. In contrast with the uniform zigzag chain, however, two eigenstates now appear inside the spectral gap.

These two states, labeled $n=20$ and $n=21$, are the only eigenstates that occupy the gap region. Their appearance demonstrates that the domain wall generates localized interface states without significantly modifying the extended bulk bands on either side of the gap.

The inverse participation ratio provides a quantitative signature of their localized character. Whereas the bulk states retain IPR values characteristic of extended eigenstates, the gap states exhibit significantly larger IPR values. The increase in the IPR relative to the uniform zigzag chain reveals a strong confinement of the wave functions around the interface.

The spatial structure of the corresponding eigenvectors is shown in Fig.~\ref{fig:DW}(d). States $n=19$ and $n=22$, which belong to the bulk bands immediately below and above the gap, remain extended throughout the system. In contrast, states $n=20$ and $n=21$ are strongly localized around the domain wall. The two localized states exhibit complementary spatial symmetries and are confined to only a few sites surrounding the interface.

The comparison between Figs.~\ref{fig:zigzag} and \ref{fig:DW} reveals an important distinction. The uniform zigzag chain develops a spectral gap but does not support localized states because the modified boundary parameters generated by the phononic mapping alter the ideal SSH picture. The introduction of a domain wall, however, restores a localized defect in the dimerization pattern while preserving the bulk spectral structure. As a consequence, a pair of localized interface states emerges inside the gap, demonstrating that geometry alone can generate SSH-like domain-wall states in the mapped phononic system.

\subsection{Transport through domain-wall states}

\begin{figure}[tb]
    \centering
    \includegraphics[width=\columnwidth]{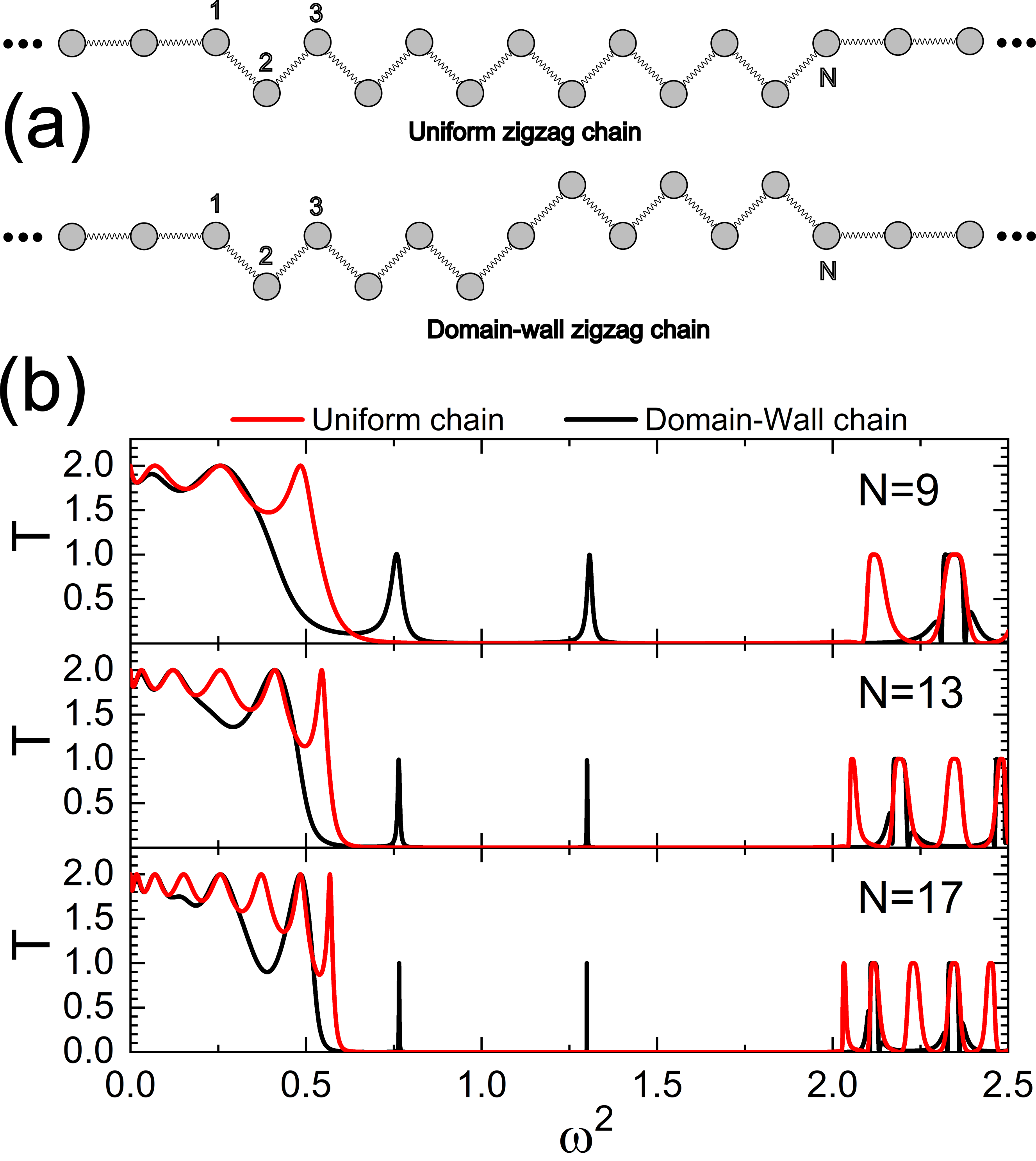}
    \caption{(a) Uniform and domain-wall zigzag chains connected to semi-infinite leads. (b) Total transmission probability as a function of frequency for different system sizes. The uniform chain exhibits a transport gap, whereas the domain-wall chain develops two resonant transmission peaks inside the gap. The resonance frequencies coincide with those of the localized domain-wall states identified in Fig.~\ref{fig:DW}.}
    \label{fig:transport}
\end{figure}

The localized states identified in the previous subsection have a direct manifestation in transport. To investigate their influence, we connect the zigzag chains to semi-infinite leads and, using the recursive S-matrix method, we calculate the total transmission probability for waves incident from the left lead,

\begin{equation}
T=T_{LL}+T_{TL}+T_{LT}+T_{TT},
\end{equation}
Figure~\ref{fig:transport}(a) compares the two geometries considered in this section: the uniform zigzag chain and the domain-wall zigzag chain. The corresponding total transmission probabilities are shown in Fig.~\ref{fig:transport}(b) for several system sizes.

In the absence of a domain wall, the transmission spectrum exhibits a broad region of vanishing transmission associated with the spectral gap identified in Fig.~\ref{fig:zigzag}(c). Waves with frequencies lying inside this interval cannot propagate through the system, resulting in a transport gap.

The situation changes qualitatively when a domain wall is introduced. Two narrow transmission resonances emerge inside the gap region, precisely at the frequencies where the localized interface states were identified in Fig.~\ref{fig:DW}(c,d). These resonances provide direct evidence that the domain-wall states act as transport channels across an energy range that remains insulating in the uniform chain.

An important feature of these resonances is their robustness against changes in system size. As the number of unit cells increases from $N=9$ to $N=17$, the resonance frequencies remain nearly unchanged, while the corresponding transmission peaks continue to approach unity. The primary effect of increasing the system size is a progressive reduction of the resonance width.

This behavior is consistent with resonant tunneling mediated by localized states concentrated around the domain wall. Because the domain-wall states are determined by the local structure of the interface, their energies are largely independent of the overall chain length. Consequently, they generate nearly perfect transmission at well-defined frequencies even though the surrounding bulk remains gapped.

The results of Fig.~\ref{fig:transport} establish a direct connection between the localized interface states identified in Fig.~\ref{fig:DW} and experimentally observable transport signatures. The geometric domain wall not only generates localized vibrational states inside the spectral gap, but also enables resonant transmission through an otherwise insulating frequency window. This demonstrates that geometry can be used to engineer both localized phononic states and frequency-selective transport channels within the mapped tight-binding framework.

\section{Conclusions}

We have developed a framework that maps harmonic vibrational systems onto effective multi-orbital tight-binding Hamiltonians. Within this construction, the Cartesian components of the displacement field become internal orbitals, while the force-constant matrix determines the effective onsite energies and hopping amplitudes. The resulting representation provides a graph-based description of phononic systems in which geometry is encoded directly in the connectivity and strength of the effective couplings.

The validity of the mapping was established through the study of a monoatomic chain containing an angular bend. For this benchmark system, an exact analytical lattice-dynamics solution was obtained and compared with calculations performed using a recursive scattering-matrix method applied to the mapped tight-binding Hamiltonian. The two approaches were found to agree to numerical precision, reproducing not only the total transmission and reflection probabilities but also all polarization-resolved scattering channels, including mode-conversion processes and the influence of evanescent states.

Having validated the mapping, we applied the framework to zigzag chains in which geometric effects generate alternating effective couplings. The mapped Hamiltonian exhibits an SSH-like structure and develops a spectral gap. Surprisingly, however, the uniform zigzag chain does not support localized edge states. We showed that this behavior originates from the modified boundary parameters generated by the phononic mapping, which alter the ideal SSH picture and suppress the expected edge-state formation.

A different behavior emerges when a geometric domain wall is introduced. In this case, a pair of localized interface states appears inside the spectral gap. These states remain localized around the domain wall and give rise to robust resonant transmission channels within an otherwise insulating frequency window. The resonance energies remain essentially independent of the system size, while the transmission approaches unity at resonance, demonstrating that the domain-wall states provide efficient pathways for phonon transport.

These results demonstrate that the mapped tight-binding description is not only a convenient reformulation of lattice dynamics, but also a powerful framework for uncovering geometry-induced transport and localization phenomena in phononic systems. More generally, the present approach enables the direct application of concepts and numerical tools originally developed for electronic transport to vibrational problems with complex geometries. The emergence of localized domain-wall states and their associated transport resonances illustrates how geometric design can be used to induce and control topological-like phononic phenomena. We expect this framework to provide a useful platform for the analysis and design of phononic structures with tailored transport and localization properties.

\begin{acknowledgement}
This work was supported by IN116025 and IA100626 UNAM-PAPIIT projects. Computations were performed at Miztli under projects LANCAD-UNAM-DGTIC-329 and  LANCAD-UNAM-DGTIC-471.
\end{acknowledgement}

\appendix

\section{Analytical solution of the angular-bend scattering problem}
\label{AppendixA}

In this appendix we present the analytical derivation of the transmission and reflection amplitudes for the monoatomic chain with an angular bend discussed in Sec.~IV.

Away from the bend, the equations of motion simplify to those of an infinite one-dimensional chain. The localized geometric mismatch affects only the vertex site ($j=0$), whose equations of motion take the form
\begin{equation}
\begin{aligned}
M\ddot{u}_{0,x}& = -\alpha u_{-1,x}+(\alpha+A)u_{0,x}+Cu_{0,y}\\
&-Au_{1,x}-Cu_{1,y} \\
 M\ddot{u}_{0,y}&=-\beta u_{-1,y}+Cu_{0,x}+(\beta+B)u_{0,y}\\
 &-Cu_{1,x}-Bu_{1,y} 
\end{aligned}
\label{motion}
\end{equation}\\
where $A=\alpha \cos^2(\theta)+\beta \sin^2(\theta)$, $B=\alpha \sin^2(\theta)+\beta \cos^2(\theta)$ and $C=(\alpha-\beta)\sin(\theta)\cos(\theta)$.

We formulate the scattering problem by considering incident waves propagating from the left ($j < 0$) toward the vertex ($j=0$). The incoming signal consists of a longitudinal (L) mode with amplitude $B_{L}^{+}$ and a transverse (T) mode with amplitude $B_{T}^{+}$. Upon scattering at the bend, these waves generate reflected longitudinal ($B_{L}^{-}$) and transverse ($B_{T}^{-}$) components propagating back into the left lead, as well as transmitted longitudinal ($A_{L}^{-}$) and transverse ($A_{T}^{-}$) components propagating along the rotated branch ($j > 0$). Indexing the vertex as $j=0$, the atomic displacement in the two semi-infinite leads are expressed as
\begin{equation}
\begin{aligned}
\mathbf{u}_j=
\frac{e^{-i\omega t}}{\sqrt{M}}
\begin{bmatrix}
\begin{pmatrix}
 B_{L}^+\\
 0\\
\end{pmatrix}e^{ijk^La}
+
\begin{pmatrix}
 B_{L}^-\\
 0\\
\end{pmatrix}e^{-ijk^La}
{~}\\
+\begin{pmatrix}
 0\\
 B_{T}^+\\
\end{pmatrix}e^{ijk^Ta}
+
\begin{pmatrix}
 0\\
 B_{T}^-\\
\end{pmatrix}e^{-ijk^Ta}
\end{bmatrix}
\end{aligned}
\label{Uj<0}
\end{equation}
for $j\leq0$, 
\begin{equation}
\mathbf{u}_j=
\frac{e^{-i\omega t}}{\sqrt{M}}
\begin{bmatrix}
A_{L}^-
\begin{pmatrix}
\cos(\theta)\\
\sin(\theta)\\
\end{pmatrix}e^{ijk^La}
{~}\\
+A_{T}^-
\begin{pmatrix}
-\sin(\theta)\\
\cos(\theta)\\
\end{pmatrix}e^{ijk^Ta}
\end{bmatrix}
\label{eq:rightlead-prop}
\end{equation}
for $j\geq0$ and 
\begin{equation}
\begin{aligned}
\mathbf{u}_0&=
\frac{e^{-i\omega t}}{\sqrt{M}}
\begin{bmatrix}
\begin{pmatrix}
A_{L}^-\cos(\theta)-A_{T}^-\sin(\theta)\\
A_{L}^-\sin(\theta)+A_{T}^-\cos(\theta)\\
\end{pmatrix}
\end{bmatrix}\\
&=
\frac{e^{-i\omega t}}{\sqrt{M}}
\begin{bmatrix}
\begin{pmatrix}
B_{L}^++B_{L}^-\\
B_{T}^++B_{T}^-\\
\end{pmatrix}
\end{bmatrix}
\end{aligned}
\label{Uj=0}
\end{equation}
for $j=0$. Here, $k^{L}$ and $k^{T}$ denote the wave vectors corresponding to the longitudinal and transverse polarizations, respectively. By substituting these displacement expressions into the vertex equations of motion (Eq.~\ref{motion}), the outgoing wave amplitudes ($A_{L}^{-}$, $A_{T}^{-}$, $B_{L}^{-}$, and $B_{T}^{-}$) are mapped linearly to the incoming amplitudes ($B_{L}^{+}$ and $B_{T}^{+}$) via the elements of the scattering matrix as functions of frequency $\omega$ and bend angle $\theta$:

\begin{equation}
   A_{L}^-=S_{LL}^tB_{L}^+ +S_{LT}^tB_{T}^+,
\end{equation}

\begin{equation}
   A_{T}^-=S_{TL}^tB_{L}^++S_{TT}^tB_{T}^+, 
\end{equation}

\begin{equation}
B_{L}^-=S_{LL}^rB_{L}^++S_{LT}^rB_{T}^+,
\end{equation}
and
\begin{equation}
B_{T}^-=S_{TL}^rB_{L}^+ +S_{TT}^rB_{T}^+,
\end{equation}
where the transmission scattering coefficients are given by
\begin{equation}
   S_{LL}^t=\frac{-2\alpha i\sin(k^La)F_4}{F_6F_4+F^2_5}, 
\end{equation}

\begin{equation}
   S_{LT}^t=\frac{2\beta i\sin(k^Ta)F_5}{F_6F_4+F^2_5},
\end{equation}

\begin{equation}
   S_{TL}^t=\frac{-2\alpha i\sin(k^La)F_5}{F_6F_4+F^2_5}, 
\end{equation}

\begin{equation}
   S_{TT}^t=\frac{-2\beta i\sin(k^Ta)F_6}{F_6F_4+F^2_5},
\end{equation}
and the reflection coefficients take the form
\begin{equation}
\begin{aligned}
S_{LL}^r=&-[F^2_3-(\alpha (X^L)^*+ F_1)(\beta X^T+F_2)]\\
&/[F^2_3-(\alpha X^L+ F_1)(\beta X^T+F_2)],
\end{aligned}   
\end{equation}

\begin{equation}
\begin{aligned}
S_{LT}^r=&[2\beta i\sin(k^Ta)F_3]\\
&/[F^2_3-(\alpha X^L+ F_1)(\beta X^T+F_2)],\\
\end{aligned}   
\end{equation}

\begin{equation}
\begin{aligned}
   S_{TL}^r=&[2\alpha i\sin(k^La)F_3]\\
   &/[F^2_3-(\alpha X^L+ F_1)(\beta X^T+F_2)],
\end{aligned}   
\end{equation}
and
\begin{equation}
\begin{aligned}
   S_{TT}^r=&-[F^2_3-(\alpha X^L+ F_1)(\beta (X^T)^*+F_2)]\\
   &/[F^2_3-(\alpha X^L+ F_1)(\beta X^T+F_2)].
\end{aligned}   
\end{equation}
The auxiliary parameters are defined as $X^L=1-e^{ik^{L}a}$, $X^T=1-e^{ik^{T}a}$, together with
\begin{equation}
\begin{aligned}
&F_1=\alpha\cos^2(\theta)X^L + \beta\sin^2(\theta)X^T - M\omega^2,\\
&F_2=\alpha\sin^2(\theta)X^L + \beta\cos^2(\theta)X^T - M\omega^2,\\
&F_3=\sin(\theta)\cos(\theta)(\alpha X^L - \beta X^T),\\
&F_4=-2i\beta\cos(\theta)\sin(k^Ta),\\
&F_5=\sin(\theta)(M\omega^2-\alpha X^L-\beta X^T),\\
&\text{and}\\
&F_6=-2i\alpha\cos(\theta)\sin(k^La).\\
\end{aligned}   
\end{equation}

The value of wave vectors $k^L$ and $k^T$ are determined directly from the bulk dispersion relations for longitudinal and transverse modes in an unperturbed infinite chain, $M\omega^2=2\alpha(1-\cos{k^La})$ and $M\omega^2=2\beta(1-\cos{k^Ta})$, respectively.

The frequency range for longitudinal propagating waves is $\omega \in [0,2\sqrt{\alpha/M}]$, while for transverse waves it is $\omega \in [0,2\sqrt{\beta/M}]$. Since typically $\beta < \alpha$, for $\omega > 2\sqrt{\beta/M}$ the transverse waves becomes evanescent. Consequently, $B_T^+ = 0$ and $T_{L,T} = T_{T,T} = R_{L,T} = R_{T,T} = 0$. Moreover, the transmitted and reflected transverse amplitudes vanish, $A_T^- = 0$ and $B_T^- = 0$, since no propagating transverse modes exist in the bent branch for $\omega > \omega_{\mathrm{max},T} = 2\sqrt{\beta/M}$. This implies that $T_{T,L} = R_{T,L} = 0$. However, the evanescent transverse modes remain coupled to the junction and therefore continue to influence the scattering process. As a result, the longitudinal transmission and reflection probabilities retain a nontrivial dependence on the transverse degree of freedom even when no transverse propagating states exist.

The frequency $\omega_{\mathrm{max},T}$, comes from $k^{T} = \pi/a$. For $\omega > \omega_{\mathrm{max},T}$, the transverse wave vector becomes complex, $k^{T} = \pi/a + iq$. This leads to the dispersion relation $M\omega^{2} = 2\beta \left[1 + \cosh(qa)\right]$. The atomic displacements for $j \leq 0$, as given in Eq.\ref{Uj<0}, are rewritten to
\begin{equation}
\begin{aligned}
\mathbf{u}_j=
\frac{e^{-i\omega t}}{\sqrt{M}}
\begin{bmatrix}
\begin{pmatrix}
 B_{L}^+\\
 0\\
\end{pmatrix}e^{ijk^La}
+
\begin{pmatrix}
 B_{L}^-\\
 0\\
\end{pmatrix}e^{-ijk^La}
{~}+\\
+
\begin{pmatrix}
 0\\
 B_{T}^-\\
\end{pmatrix}(-1)^je^{jqa}.
\end{bmatrix}
\end{aligned}
\end{equation}
Analogously, for $j\ge 0$ the Eq. \ref{eq:rightlead-prop} becomes
\begin{equation}
\mathbf{u}_j=
\frac{e^{-i\omega t}}{\sqrt{M}}
\begin{bmatrix}
A_{L}^-
\begin{pmatrix}
\cos(\theta)\\
\sin(\theta)\\
\end{pmatrix}e^{ijk^La}
{~}+\\
A_{T}^-
\begin{pmatrix}
-\sin(\theta)\\
\cos(\theta)\\
\end{pmatrix}(-1)^je^{-jqa}.
\end{bmatrix}
\label{eq:rightlead-ev}
\end{equation}

Consequently, for $j=0$, in the atomic displacements given by Eq. \ref{Uj=0}, the only change is $B_T^+=0$. Substituting these atomic displacements into the motion Eq.\ref{motion}, we obtain new expressions for $B_L^-$ and $A_L^-$, which are given by
\begin{equation}
   A_{L}^-=\tilde{S}_{LL}^tB_{L}^+,
\end{equation}
and
\begin{equation}
B_{L}^-=\tilde{S}_{LL}^rB_{L}^+
\end{equation}
being
\begin{equation}
   \tilde{S}_{LL}^t=\frac{-2\alpha i\sin(k^La)\tilde{F}_4}{\tilde{F}_6\tilde{F}_4+\tilde{F}^2_5},  
\end{equation}
and 
\begin{equation}
\begin{aligned}
&\tilde{S}_{LL}^r=\\
&[\tilde{F}^2_3-(\alpha (X^L)^*+ \tilde{F}_1)(\beta(1+e^{-qa})+\tilde{F}_2)]\\
&/[\tilde{F}^2_3-(\alpha X^L+ \tilde{F}_1)(\beta(1+e^{-qa})+\tilde{F}_2)],
\end{aligned}   
\end{equation}
where
\begin{equation}
\begin{aligned}
&\tilde{F}_1=\alpha \cos^2(\theta)X^L+\beta\sin^2(\theta)(1+e^{-qa})-M\omega^2,\\
&\tilde{F}_2=\alpha \sin^2(\theta)X^L+\beta\cos^2(\theta)(1+e^{-qa})-M\omega^2,\\
&\tilde{F}_3=\sin(\theta)\cos(\theta)(\alpha X^L-\beta(1+e^{-qa})),\\
&\tilde{F}_4=\cos(\theta)(2\beta(1+e^{-qa})-M\omega^2),\\
&\tilde{F}_5=\sin(\theta)(M\omega^2-\alpha X^L- \beta(1+e^{-qa})),\\
&\text{and}\\
&\tilde{F}_6=\cos(\theta)(2\alpha X^L-M\omega^2)\\
\end{aligned}   
\end{equation}
In this scenario, $T_{L,L}=|\tilde{S}_{LL}^t|^2$ and $R_{L,L}=|\tilde{S}_{LL}^r|^2$.

\section{Analytical Proof of $T_{LL} = T_{TT}$ and $T_{TL} = T_{LT}$}
\label{AppendixB}
\subsection{Proof of $T_{LL} = T_{TT}$}
To prove that $T_{LL} = T_{TT}$, it suffices to show that $A^{LL}$ and $A^{TT}$ are equal, since they represent the numerators of $S^t_{LL}$ and $S^t_{TT}$, which share the same denominator. Given:
\begin{equation}
A^{LL} = -2i\alpha\sin(k^L a) F_4
\end{equation}
and
\begin{equation}
A^{TT} = -2i\beta\sin(k^T a) F_6,
\end{equation}
substituting $F_4 = -2i\beta\cos(\theta)\sin(k^T a)$ into $A^{LL}$ and $F_6 = -2i\alpha\cos(\theta)\sin(k^L a)$ into $A^{TT}$, both amplitudes reduce to the exact same expression:
\begin{equation}
A^{LL} = A^{TT} = -4\alpha\beta\cos(\theta)\sin(k^L a)\sin(k^T a).
\end{equation}
Therefore,

\[
T_{LL}=T_{TT}.
\]

\subsection{Proof of $T_{LT} = T_{TL}$}

Since $T_{TL} = \frac{v_T}{v_L} |S_{TL}^t|^2$ and $T_{LT} = \frac{v_L}{v_T} |S_{LT}^t|^2$, it suffices to show that:
\begin{equation}
v_L^2 |A^{LT}|^2 = v_T^2 |A^{TL}|^2,
\end{equation}
where
\begin{equation}
A^{LT} = 2\beta i \sin(k^T a) F_5
\end{equation}
and
\begin{equation}
A^{TL} = -2\alpha i \sin(k^L a) F_5.
\end{equation}

First, we calculate the squared moduli of the amplitudes:
\begin{equation}
|A^{LT}|^2 = 4\beta^2 \sin^2(k^T a) |F_5|^2
\end{equation}
and
\begin{equation}
|A^{TL}|^2 = 4\alpha^2 \sin^2(k^L a) |F_5|^2.
\end{equation}

Using the double-angle identity for the sine ($\sin(2x) = 2\sin(x)\cos(x)$), we expand the squared sines in each expression:
\begin{align}
|A^{LT}|^2 &= 16\beta^2 \sin^2\left(\frac{k^T a}{2}\right) \cos^2\left(\frac{k^T a}{2}\right)|F_5|^2 \\
|A^{TL}|^2 &= 16\alpha^2 \sin^2\left(\frac{k^L a}{2}\right) \cos^2\left(\frac{k^L a}{2}\right)|F_5|^2
\end{align}

On the other hand, the squared group velocities can be expressed as:
\begin{align}
v_L^2 &= a^2 \frac{\alpha}{M} \cos^2\left(\frac{k^L a}{2}\right) \\
v_T^2 &= a^2 \frac{\beta}{M} \cos^2\left(\frac{k^T a}{2}\right).
\end{align}

Then,
\begin{equation}
\begin{aligned}
v_L^2 |A^{LT}|^2 &= a^2 4\alpha\beta \cos^2\left(\frac{k^L a}{2}\right)\cos^2\left(\frac{k^T a}{2}\right)|F_5|^2 \\
&\quad \times \frac{4\beta}{M}\sin^2\left(\frac{k^T a}{2}\right) \\
v_T^2 |A^{TL}|^2 &= a^2 4\alpha\beta \cos^2\left(\frac{k^T a}{2}\right)\cos^2\left(\frac{k^L a}{2}\right)|F_5|^2 \\
&\quad \times \frac{4\alpha}{M}\sin^2\left(\frac{k^L a}{2}\right).
\end{aligned}
\end{equation}

From the dispersion relation, we can equate the expressions for the eigenfrequencies of the longitudinal and transverse modes:
\begin{equation}
\omega^2 = \frac{4\alpha}{M}\sin^2\left(\frac{k^L a}{2}\right) = \frac{4\beta}{M}\sin^2\left(\frac{k^T a}{2}\right).
\end{equation}

Applying this equivalence, both expressions become identical; thus,
\begin{equation}
v_L^2 |A^{LT}|^2 = v_T^2 |A^{TL}|^2.
\end{equation}

Therefore,

\[
T_{TL}=T_{LT}.
\]

\bibliography{references.bib}

\end{document}